\documentclass[largeformat]{interact}

\usepackage{epstopdf}
\usepackage[caption=false]{subfig}
\usepackage{amsmath}
\mathchardef\mhyph="2D 
\usepackage[usenames,dvipsnames]{xcolor}
\usepackage{hyperref}
\usepackage[numbers,sort&compress,merge]{natbib}
\bibpunct[, ]{[}{]}{,}{n}{,}{,}

\theoremstyle{plain}

\theoremstyle{definition}

\theoremstyle{remark}

\definecolor{BlueF}{rgb}{0.00, 0.53, 0.74}

\newcommand\apj{Astrophys. J. } 
\newcommand\apjs{Astrophys. J. Suppl. Ser. } 
\newcommand\mnras{Monthl. Not. R. Astr. Soc. } 
\begin{document}

\articletype{RESEARCH ARTICLE}

\title{Understanding the temperatures of $\mathrm{H}_3^+$  and $\mathrm{H}_2$ in diffuse interstellar sightlines}

\author{
\name{Jacques Le Bourlot\textsuperscript{a,c}\thanks{Email: Jacques.Lebourlot@obspm.fr}, Evelyne Roueff\textsuperscript{\,a}, Franck Le Petit\textsuperscript{a},\\
Florian Kehrein\textsuperscript{b}, Annika Oetjens\textsuperscript{b}, and Holger Kreckel\textsuperscript{b}\thanks{Email: holger.kreckel@mpi-hd.mpg.de} }
\affil{\textsuperscript{a}LERMA, Observatoire de Paris, PSL Research University, CNRS, Sorbonne Universités, 92190 Meudon, France; \textsuperscript{b}Max-Planck-Institut f\"ur Kernphysik, 69117 Heidelberg, Germany; \\ \textsuperscript{c}Université Paris Cité
}}

\maketitle

\begin{abstract}
The triatomic hydrogen ion $\mathrm{H}_3^+$ is one of the most important species for the gas phase chemistry of the interstellar medium. Observations of $\mathrm{H}_3^+$ are used to constrain important physical and chemical parameters of interstellar environments. However, the temperatures inferred from the two lowest rotational states of $\mathrm{H}_3^+$ in diffuse lines of sight -- typically the only ones observable --  appear consistently lower than the temperatures derived from H$_2$ observations in the same sightlines. All previous attempts at modeling the temperatures of $\mathrm{H}_3^+$ in the diffuse interstellar medium failed to reproduce the observational results.

Here we present new studies, comparing an independent master equation for H$_3^+$ level populations to results from the Meudon PDR code for photon dominated regions. We show that the populations of the lowest rotational states of $\mathrm{H}_3^+$ are strongly affected by the formation reaction and that H$_3^+$ ions experience incomplete thermalization before their destruction by free electrons. Furthermore, we find that for quantitative analysis more than two levels of H$_3^+$ have to be considered and that it is crucial to include radiative transitions as well as collisions with H$_2$. Our models of typical diffuse interstellar sightlines show very good agreement with observational data, and thus they may finally resolve the perceived temperature difference attributed to these two fundamental species. 

\end{abstract}



\section{Introduction} \label{sec:intro}

The triatomic hydrogen ion $\mathrm{H}_3^+$ is one of the main drivers of interstellar chemistry in the gas phase \cite{oka2006}. It is formed very efficiently in the interstellar medium by collisions between hydrogen molecules and hydrogen molecular ions
\begin{equation}
\label{eq:h3formation}
{\rm H}_2 + {\rm H}_2^+ \longrightarrow \mathrm{H}_3^+ + {\rm H}\,.
\end{equation}
Owing to the comparatively low proton affinity of $\mathrm{H}_2$, triatomic hydrogen readily reacts with many of the neutral atomic and molecular species present in interstellar environments. By donating a proton in exothermic ion-neutral collisions of the type
\begin{equation}
    \mathrm{H}_3^+ + \mathrm{X} \longrightarrow \mathrm{XH}^+ + \mathrm{H}_2\,,
\end{equation}
triatomic hydrogen often initiates gateway processes, which subsequently enable the formation of more complex molecules in interstellar space (here the X stands for any neutral atomic or molecular collision partner). In particular, reactions between $\mathrm{H}_3^+$ ions and neutral $\mathrm{O}$, and $\mathrm{C}$ 
atoms will lead to the formation of $\mathrm{OH}^+$,   and $\mathrm{CH}^+$ 
respectively, and thus facilitate the introduction of the heavier atomic species into the chemical networks. 

The relevance of $\mathrm{H}_3^+$ for interstellar chemistry was recognized already in early quantitative models of interstellar clouds \cite{watson1973, herbst1973}. After the breakthrough work of Oka \cite{oka1980}, who identified the infrared spectrum of the $\mathrm{H}_3^+$ fundamental vibrational band in the laboratory, $\mathrm{H}_3^+$ was found in both dense \cite{geballe1996} and diffuse lines of sight \cite{mccall1998}, confirming the role of ion-neutral chemistry in space. 

In the meantime, triatomic hydrogen has been detected in various interstellar sightlines, including the Galactic center, as well as extra-galactic sources and planetary atmospheres (see \cite{miller2020} for a recent review of $\mathrm{H}_3^+$ astronomy). Moreover, owing to its seemingly simple formation and destruction mechanisms, $\mathrm{H}_3^+$ observations have been used to constrain important astrophysical parameters like, e.g., the cosmic ray ionization rate \cite{mccall2003, indriolo2010, indriolo2012, lepetit2016} and the temperatures and densities of the molecular gas in the vicinity of the Galactic center \cite{oka2019}. Submillimeter observations of H$_2$D$^+$, an isotopic variant of triatomic hydrogen, have been used to infer the minimum age of a star-forming molecular cloud \cite{bruenken2014}.  

While the hydrogen molecule $\mathrm{H}_2$ is by far the most abundant molecule in space, the bulk of $\mathrm{H}_2$ molecules in colder environments is difficult to observe with ground-based telescopes. Direct observation of $\mathrm{H}_2$ in diffuse and translucent interstellar clouds are obtained either from satellite absorption observations of electronic transitions in the ultraviolet regime (see, e.g., \cite{spitzer1974, shull2021}) towards bright stellar sources or
by infra-red quadrupolar electric emission of its rovibrational spectrum in bright and dense photodissociation regions (PDRs). 

However, attempts to understand the observed column densities $N_0$ and $N_1$ of the two lowest rotational states of $\mathrm{H}_2$ (with $J=0$ and $J=1$, respectively) and the column densities $N_{(1,1)}$ and $N_{(1,0)}$ of the two lowest states of $\mathrm{H}_3^+$ (with $(J,G)=(1,1)$ and $(J,G)=(1,0)$, respectively) in the same lines of sight led to a surprise \cite{crabtree2011}. The temperature  derived from the lowest states of $\mathrm{H}_3^+$, 
$T_{12}\left(\mathrm{H}_3^+\right) = 32.9 / \ln (2\, N_{(1,1)}/ N_{(1,0)})\,\mathrm{K}$ 
appeared systematically lower than the temperature of the lowest states of $\mathrm{H}_2$, $T_{01} = 170.5 / \ln(9\, N_0/ N_1)\,\mathrm{K}$, which is usually found to be in equilibrium with the gas kinetic temperature \cite{crabtree2011, shull2021}. While the initial studies used very simple model calculations for the $\mathrm{H}_3^+$ abundances and thermalization processes, later models, employing large chemical networks,  focused principally on the chemical evolution of the ortho/para forms of $\mathrm{H}_2$ and $\mathrm{H}_3^+$ abundances\footnote{In fact, only the lowest {\em para} and {\em ortho} levels are considered in these studies.} without considering detailed collisional excitation  mechanisms nor introducing thermal balance considerations \cite{albertsson2014}.      

Here, we present a different approach and introduce first a master equation describing the evolution of rotational levels of $\mathrm{H}_3^+$, based on updated rate coefficients for collisional and chemical processes at fixed density and temperature, which can be solved for steady state at definite physical conditions. 
This procedure is able to reproduce the observational trends and temperature differences on a quantitative level, and
it allows for an analysis of the contributions of the various processes.
We further introduce  the same mechanisms in our Meudon PDR code \cite{lepetit2006}, which solves both the chemical and thermal equilibrium of the cloud and add their contribution to the equilibrium state of $\mathrm{H}_2$, where photodissociation, collisional excitation and chemical formation/destruction mechanisms are considered together.

The paper is organized as follows: Section \ref{sec:nucspin} presents a brief overview of nuclear spin and the rotational levels of $\mathrm{H}_3^+$ and $\mathrm{H}_2$. In Section\,\ref{sec:proc} we describe the most important processes driving the {\em ortho-para} ratio of $\mathrm{H}_3^+$. A master equation for H$_3^+$ state populations is presented in Section\,\ref{sec:detailed_balance} together with the corresponding results on the excitation temperature of $\mathrm{H}_3^+$. We compare our model results to astronomical observations in Section \ref{sec:results}, where we also introduce the modifications
included in the PDR code to account for the {\em ortho-para} character of $\mathrm{H}_3^+$. The paper concludes with a brief discussion in Section\,\ref{sec:discussion}.

\section{Nuclear spin and rotational states of \texorpdfstring{$\mathrm{H}_3^+$}{H3+} and \texorpdfstring{$\mathrm{H}_2$}{H2} } \label{sec:nucspin}

\begin{figure}
\begin{center}
\includegraphics[width=0.95\textwidth]{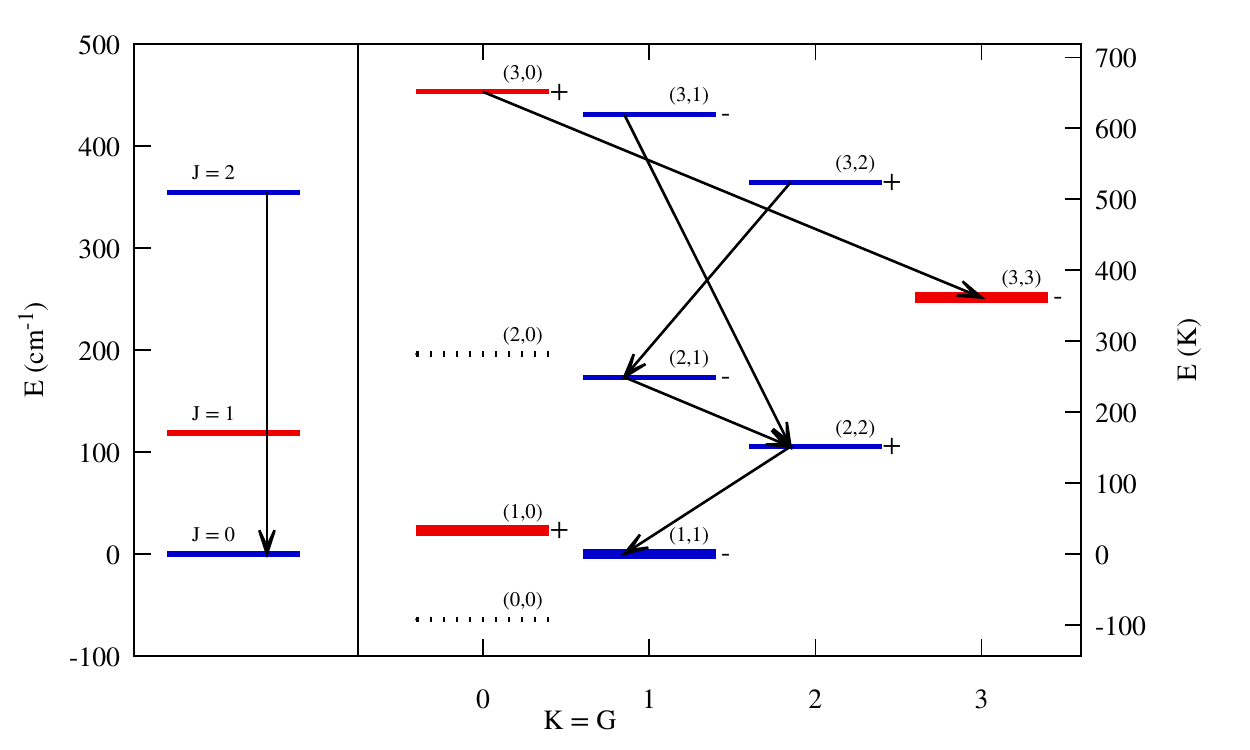}
\end{center}
\caption{Level scheme of all rotational levels with $J \le 2$ for $\mathrm{H}_2$ and with $J \le 3$ for $\mathrm{H}_3^+$.
The level energy (in $\mathrm{cm}^{-1}$) is given on the y-axis at the left-hand-side, while the corresponding values in $\mathrm{K}$ are given on the right-hand-side of the graph.
{\em Para} levels are in blue and {\em ortho} levels in red. Black arrows show the possible radiative transitions. Stable and metastable levels are shown with a heavier line.
The symmetry-forbidden levels $(0,0)$ and $(2,0)$ are shown by dotted lines for completeness. The origin of the energies 
is fixed at the lowest permitted level. \label{fig1}}
\end{figure}

Both $\mathrm{H}_2$ and $\mathrm{H}_3^+$ exist in two different nuclear spin configurations. For the lowest rotational state of $\mathrm{H}_2$, with $J=0$, the proton spins are anti-parallel and add up to $I=0$. This configuration is denoted as $para\mhyph\mathrm{H}_2$ (or $p\mhyph\mathrm{H}_2$). The next highest level is $J=1$, and, owing to the requirement that the total wave function has to change sign under the permutation of both protons, all $\mathrm{H}_2$ states with odd rotational quantum numbers have parallel nuclear spin configurations with $I=1$, which is denoted as $ortho\mhyph\mathrm{H}_2$ (or $o\mhyph\mathrm{H}_2$). Likewise, all even rotational quantum numbers in H$_2$ can be attributed to $p\mhyph\mathrm{H}_2$ with $I=0$. Figure\,\ref{fig1} shows the three lowest levels of $\mathrm{H}_2$ (with $J\le 2$) and their respective energies expressed as $k_b T$ (in units of kelvin).     

The excitation temperature of the two lowest states of $\mathrm{H}_2$ is defined as
\begin{equation}
    T_{01} = \frac{\Delta E_{01}/ k_b} {\ln( g_1/g_0 \,\cdot N_0/N_1)}\,,
\end{equation}
where $\Delta E_{01}/ k_b = 170.476\,\mathrm{K}$ stands for the energy difference between the states, $g_1/g_0=9$ is the ratio of the multiplicities of the states with $J=1$ and $J=0$, respectively, and $N_1$ and $N_0$ denote their populations. An analogous equation can be given for the excitation temperature $T_{02}$ between   the states with $J=2$ and $J=0$, for which $\Delta E_{02}/ k_b =509.864$\,K and $g_2/g_0=5$,  and both temperatures are usually found to be consistent with one another \cite{shull2021}, giving a good proxy of the kinetic gas temperature.
Typical values for $T_{01}$ in these diffuse sightlines range from $50$ to $70\,\mathrm{K}$ \cite{savage1977, rachford2002, rachford2009, shull2021}.  

The nuclear spin configurations of $\mathrm{H}_3^+$ are also denoted by \emph{para} and \emph{ortho} for $I=1/2$ and $I=3/2$, respectively. As with $\mathrm{H}_2$, the symmetry of the nuclear spin wave function imposes restrictions on the rotational quantum numbers. The relevant quantum number is usually denoted by $G$, which is connected to the projection of the angular momentum (from both vibrational and rotational motion) onto the molecular symmetry axis (see the review by \cite{lindsay2001} for quantum numbers, symmetries and selection rules). Since we are only concerned with molecules in the vibrational ground state here, the quantum number $G$ can be regarded as equivalent with the projection $K$ of the rotational angular momentum (denoted by quantum number $J$) onto the normal axis of the molecular plane, implying $G=K$. The symmetry of the total wave function requires $G=3n$ (where $n$ is an integer) for the $I=3/2$ or \emph{ortho} levels of $\mathrm{H}_3^+$, while all other levels (effectively fulfilling $G=3n \pm 1$) are of the \emph{para} configuration, with $I=1/2$. The right-hand side of Figure\,\ref{fig1} shows all $\mathrm{H}_3^+$ rotational levels with $J \le 3$ and their respective energies. Because of the Pauli exclusion principle, pure rotational levels with even $J$ and $G=0$ do not exist. This rules out the nominal $(J,G)=(0,0)$ ground state and the $(J,G)=(2,0)$ state, which are both indicated in the graph by dotted lines for completeness. Since we are almost exclusively concerned with rotational states in the vibrational ground state of $\mathrm{H}_3^+$, we will from now on refer to all $\mathrm{H}_3^+$ states by giving their $(J,G)$ quantum numbers only, e.g., $(1,1)$ and $(1,0)$ for the lowest {\em para} and {\em ortho} states, respectively. 

With the notable exception of the Galactic center \cite{goto2002,oka2019, oka2022}, only the lowest $(1,1)$ {\em para}-state of $\mathrm{H}_3^+$  and the lowest \emph{ortho}-state $(1,0)$ have ever been detected in the interstellar medium. Consequently, the $\mathrm{H}_3^+$ excitation temperature, $T_{12}(\mathrm{H}_3^+)$, derived from observations is usually calculated using the column density ratio of these two states 
\begin{equation}
\label{T01H3p}
 T_{12}(\mathrm{H}_3^+) = \frac{\Delta E / k_b} {\ln(( g_{1,0}/g_{1,1}) \,\cdot N_{(1,1)}/N_{(1,0)})}\,,
\end{equation}
where $\Delta E/ k_b = 32.86\,\mathrm{K}$, and 
$g_{1,0}=12$ and $g_{1,1}=6$ denote the total degeneracies of the \emph{ortho}- and \emph{para}-state, respectively. Most observations yield values for $T_{12}(\mathrm{H}_3^+)$ between $20$ and $40\,\mathrm{K}$ \cite{crabtree2011,albertsson2014}, systematically lower than the $\mathrm{H}_2$ excitation temperature $T_{01}$.   
As \emph{para}- and \emph{ortho}-states of $\mathrm{H}_2$ and $\mathrm{H}_3^+$ can not be inter-converted by radiative transitions, we now describe the 
various collisional and reactive processes allowing \emph{ortho}/\emph{para} exchange.

\section{Processes controlling the para fraction of \texorpdfstring{$\mathrm{H}_3^+$}{H3+}} \label{sec:proc}

\subsection{Formation of \texorpdfstring{$\mathrm{H}_3^+$}{H3+}}
$\mathrm{H}_3^+$ formation via the $\mathrm{H}_2^+$ + $\mathrm{H}_2$ reaction (Eq.\,\ref{eq:h3formation}) has been studied for many years by various experimental techniques. A recent compilation of the results can be found in \cite{savic2020}. Particularly noteworthy is a recent study that employs excited Rydberg molecules in a merged beams approach to reach collision energies between $5-60\,\mathrm{K}$ \cite{allmendinger2016}. The results are in very good agreement with the previous measurements \cite{glenewinkel1997} conducted at somewhat higher energies. A recommended fit to the overall cross section can be found in \cite{savic2020}, yielding values at low temperature that are slightly higher than the corresponding values derived from the classical Langevin collision rate. We have converted this cross section to a thermal rate coefficient
\begin{equation} \label{k_form}
   k_{\rm form}= 2.27\,\,10^{-9} \times \left(\frac{T}{300}\right)^{-0.06} \quad  {\rm cm}^{3}\,{\rm s}^{-1} \, ,
\end{equation}
with $T$, the gas kinetic temperature, in kelvin. This rate shows only a weak dependence on temperature. Its value is slightly larger than the constant rate of $2.08\times 10^9\,\mathrm{cm}^3\,\mathrm{s}^{-1}$ that is reported in most astrochemistry databases (and based on a previous experimental study \cite{theard1974}, which was conducted at room temperature). 

To determine the nuclear spin of the $\mathrm{H}_3^+$ ions created by reaction\,(\ref{eq:h3formation}), we refer to the selection rules outlined by Oka \cite{oka2004}. While this scheme is based on pure angular momentum algebra, and thus does not take any energetic barriers or other restrictions of the reaction into account, we consider this to be a very good approximation for the highly exothermic barrier-less ion neutral reaction between $\mathrm{H}_2^+$ and $\mathrm{H}_2$. To quantify the outcome of the reaction in terms of nuclear spin, it is convenient to use the {\em para}-fractions of both molecular species derived from the densities $n(p\mhyph\mathrm{H}_2)$ and $n(o\mhyph\mathrm{H}_2)$ of the \emph{para}- and \emph{ortho}-species of $\mathrm{H}_2$, and  $n(p\mhyph\mathrm{H}_3^+)$ and $n(o\mhyph\mathrm{H}_3^+)$ of $\mathrm{H}_3^+$, respectively. The {\em para}-fraction for both species is then denoted by
\begin{equation}
    p_2 = \frac{n(p\mhyph\mathrm{H}_2)}{n(p\mhyph\mathrm{H}_2) + n(o\mhyph\mathrm{H}_2)}
\end{equation}
and 
\begin{equation} \label{p3}
    p_3 = \frac{n(p\mhyph\mathrm{H}_3^+)}{n(p\mhyph\mathrm{H}_3^+) + n(o\mhyph\mathrm{H}_3^+)}\,.
\end{equation}
Following the argumentation by Crabtree et al. \cite{crabtree2011}, we assume that the cosmic ray ionization of $\mathrm{H}_2$ (constituting the main source of ionization that initiates the $\mathrm{H}_3^+$ formation) does not affect the nuclear spin of the molecule, and therefore the \emph{para}-fraction of $\mathrm{H}_2^+$ has the same value $p_2$. 

\begin{figure}
\begin{center}
\includegraphics[width=0.95\textwidth]{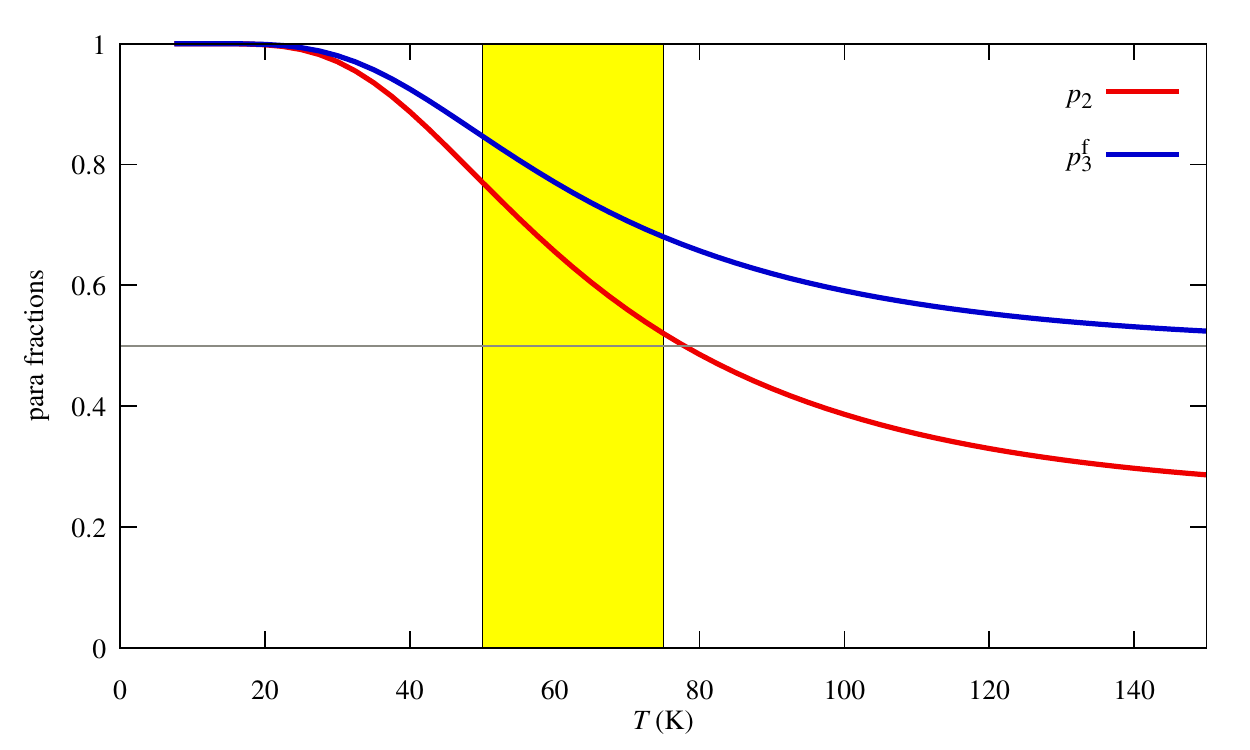}
\end{center}
\caption{Fraction $p_2$ of $p\mhyph\mathrm{H}_2$ and $p_3^f$ of nascent $p\mhyph\mathrm{H}_3^+$ (Eq.\,\ref{f3}). The typical range of diffuse cloud kinetic temperature is outlined in yellow. \label{fig2}}
\end{figure}

With these assumptions the $p\mhyph\mathrm{H}_3^+$ fraction at formation, $p_3^f$, can be derived as a function of  $p_2$, using the nuclear spin branching fractions given in \cite{oka2004}  (for details see Table\,4 in \cite{crabtree2011}), resulting in a simple linear dependence of the form
\begin{equation} \label{f3}
    p_3^f = \frac{1+2\,p_2}{3} \,.
\end{equation}
The values of $p_2$ and $p_3^f$ 
are displayed as a function of temperature in Figure\,\ref{fig2}  for $\mathrm{H}_2$ in thermal equilibrium.

We further assume that the rotational levels of $p\mhyph\mathrm{H}_3^+$ and $o\mhyph\mathrm{H}_3^+$ are each populated at formation according to their Boltzmann distribution at a temperature corresponding to $2/3$ of the reaction exothermicity (\cite{merkt2022}). Given the low energy of the relevant levels, this amounts in effect at using rates proportional to the statistical weight of the level. 

\subsection{Thermalizing collisions of \texorpdfstring{$\mathrm{H}_3^+$}{H3+}}
Collisions with electrons, $\mathrm{He}$, $\mathrm{H}$ and $\mathrm{H}_2$ contribute to the energy exchange between the  rovibrational levels of $\mathrm{H}_3^+$, but only collisions with $\mathrm{H}_2$ and $\mathrm{H}$ may change the nuclear spin of the $\mathrm{H}_3^+$ ions. To the best of our knowledge, of the above collision partners, only for collisions with $\mathrm{H}_2$ and electrons detailed information is available in the literature.

\subsubsection{Collisions between \texorpdfstring{$\mathrm{H}_3^+$}{H3+} and \texorpdfstring{$\mathrm{H}_2$}{H2}}
Before any detailed study of $\mathrm{H}_3^+$ collision rates with $\mathrm{H}_2$ was available, Oka and Epp \cite{oka2004b}  suggested to use the Langevin expressions to describe the excitation of $\mathrm{H}_3^+$ detected towards the Galactic center. 
However, as a result of the five identical Fermion nuclei involved in these collisions, \cite{oka2004,park2007,hugo2009} pointed out that 
one should consider the nuclear spin dependence of the total wavefunction.
Three different possibilities have to be accounted for
\begin{subequations}
\begin{eqnarray}
\hspace*{-1cm} \mathrm{H}_3^+ + \tilde{\mathrm{H}}_2 &~  \longrightarrow ~ & \mathrm{H}_3^+ +\tilde{\mathrm{H}}_2  \hspace*{1.35cm}{\rm identity} \hspace*{1.60cm} \sim {\rm inelastic\,\,collision},  \\
                                 & \longrightarrow & \mathrm{H}_2 + (\mathrm{H}\tilde{\mathrm{H}}_2)^+  \hspace*{.65cm}{\rm proton-hop} \hspace*{.65cm}\sim{\rm reactive\,\,collision},  \\
                                 & \longrightarrow & \mathrm{H}\tilde{\mathrm{H}} +(\tilde{\mathrm{H}}\mathrm{H}_2)^+ \hspace*{0.55cm}{\rm exchange}  \hspace*{1.35cm}\sim{\rm reactive\,\,collision} . 
\end{eqnarray}
\end{subequations}
Specific selection rules on the nuclear spins of the products can be derived, following \cite{quack1977,oka2004}, and they have been used to model hydrogen plasma experiments \cite{crabtree2011}. The calculations of \cite{park2007} compared favourably to ion trap measurements of the nuclear spin equilibrium of $\mathrm{H}_3^+$ and $\mathrm{H}_2$ at low temperature \cite{grussie2012}, although these studies did not allow for detailed comparisons of the absolute rate coefficients.

For our models, we have adopted the rate coefficients resulting from the most advanced theoretical treatment of the $\mathrm{H}_3^+$ -- $\mathrm{H}_2$ reaction so far, which was presented by \cite{gomez2012}. Their approach is similar to the method of \cite{park2007}, but refined by a dynamical bias that is introduced through a scrambling matrix, accounting for the relative probabilities of the identity/hop/exchange channels. The probabilities are calculated using quasi-classical trajectory calculations, based on a global $\mathrm{H}_5^+$ potential energy surface \cite{aguado2010}. 
We employ a set of rate coefficients that covers collisions of the lowest $24$ rotational states of $\mathrm{H}_3^+$ with {\em ortho}-$\mathrm{H}_2$ and {\em para}-$\mathrm{H}_2$ (in their lowest rotational states) and temperatures up to $500\,\mathrm{K}$, which were kindly provided by O. Roncero. Those rate coefficients are also currently used in our PDR model calculations \cite{lepetit2006}.

\subsubsection{Collisions between {\texorpdfstring{$\mathrm{H}_3^+$}{H3+}},    
{\texorpdfstring{$\mathrm{He}$}{He}} and {\texorpdfstring{$\mathrm{H}$}{H}}}
$\mathrm{He}$ and $\mathrm{H}$ are additional collision partners that should be taken into account when describing the excitation equilibrium of $\mathrm{H}_3^+$, but, to the best of our knowledge, no detailed studies are
presently available on these systems. Collisions of $\mathrm{H}_3^+$ with $\mathrm{He}$ can not modify the nuclear spin configuration of $\mathrm{H}_3^+$, and we approximate the corresponding collision rates by taking the rates for $p\mhyph\mathrm{H}_2$ and scaling them with the reduced mass factor (while vetoing channels that might change the nuclear spin of $\mathrm{H}_3^+$), as described previously \cite{roueff2013}.

Collisions of $\mathrm{H}$ with $\mathrm{H}_3^+$ may lead to proton exchange or inelastic collisions, similar to the channels discussed for the $\mathrm{H}_3^+ + \mathrm{H}_2$ reaction. However, since we are not aware of more detailed information on this process, we will use the collisional rates with $p\mhyph\mathrm{H}_2$ as a proxy for collisions with $\mathrm{H}$. We checked that this choice has no major influence on our model calculations for the conditions studied here.

\subsubsection{Inelastic collisions between \texorpdfstring{$\mathrm{H}_3^+$}{H3+} and electrons}

Electronic collisions with $\mathrm{H}_3^+$ preserve the \emph{ortho}-\emph{para} character of $\mathrm{H}_3^+$.  They have been computed by \cite{kokoouline2010} and are included in the present model. Until very recently no laboratory measurements of the change of rotational states in electron collisions were available for any molecular ion to benchmark the theoretical approach. But a recent measurement of low-energy electron collisions with $\mathrm{CH}^+$ allowed for a comparison between experiment and theory for the lowest rotational states \cite{kalosi2022}, which revealed very good agreement. 

\subsection{Dissociative recombination of \texorpdfstring{$\mathrm{H}_3^+$}{H3+}} \label{subsec:Recomb_elec}
In diffuse gas, the main destruction reaction of $\mathrm{H}_3^+$ is the dissociative recombination (DR) with  electrons.
\begin{subequations}
\begin{eqnarray}
\hspace*{-1cm} \mathrm{H}_3^+ + e^- & \longrightarrow & \mathrm{H} + \mathrm{H} + \mathrm{H} \label{11a}, \\
                                     & \longrightarrow & \mathrm{H}_2 + \mathrm{H} \label{11b}.
\end{eqnarray}\nonumber
\end{subequations}
This is an important chemical reaction for interstellar chemistry, and as such, has attracted a lot of attention, as well as some controversy. More than 30 independent experimental studies of the DR rate coefficient of $\mathrm{H}_3^+$ have been published, with outcomes that differ by orders of magnitude (there are a number of reviews on this topic, see, e.g, \cite{larsson2008,larsson2012,kreckel2012}).
In summary, the present consensus is that the absolute rate of the $\mathrm{H}_3^+$ DR rate coefficient is correctly derived from the storage ring merged beams measurements \cite{kreckel2012}. Theoretical studies identified the Jahn-Teller effect as a driver for the recombination process at low temperatures \cite{kokoouline2001,kokoouline2003,dossantos2007}.

Both astrochemistry databases, UMIST\footnote{available at
\href{http://www.udfa.net}{http://www.udfa.net}} and KIDA\footnote{available at
\href{https://kida.astrochem-tools.org}{https://kida.astrochem-tools.org}}, report rate coefficients  with branching ratios of $2/3$ for reaction\,(\ref{11a}) and $1/3$ for reaction\,(\ref{11b}), based on storage ring experiments, and the total DR reaction rate coefficient is given as

\begin{equation}
\alpha_{\rm DR}^{tot}= 6.70 \times 10^{-8} \, \left(\frac{T}{300}\right)^{-0.52}\,\mathrm{cm}^3 \, \mathrm{s}^{-1}\,.
\end{equation}
However, there are recent suggestions concerning a possible difference between the DR rate coefficients of the two nuclear spin modifications of $\mathrm{H}_3^+$. These were first pointed out in the storage ring experiments of \cite{kreckel2005,kreckel2010}, but the values are dependent on the actual rotational level populations, which could not be determined precisely. 
 Subsequent plasma experiments found an even stronger effect at low temperature \cite{glosik2011}, and updated theoretical studies \cite{dossantos2007} predict that at low temperature the difference between the rate coefficients for the two nuclear spin modifications may exceed  an order of magnitude, with $p\mhyph\mathrm{H}_3^+$ recombining much faster than $o\mhyph\mathrm{H}_3^+$.  
 Two different theoretical values  are explicitly reported in \cite{pagani2009}, supporting plasma studies. We 
 have derived an analytic expression from these results, which we can describe by the formulae 
\begin{subequations}
\begin{eqnarray}
\alpha_{\rm DR}^{tot}(p\mhyph\mathrm{H}_3^+) &=& 5.25 \times 10^{-8} (T/300)^{-0.75}\,{\rm cm}^3 \,{\rm s}^{-1}, \label{alp_DR_p} \\[0.3cm]
\alpha_{\rm DR}^{tot}(o\mhyph\mathrm{H}_3^+) &=& 6 \times 10^{-8}     \,{\rm cm}^3\, {\rm s}^{-1} \qquad T \leq 250\,{\rm K} \label{alp_DR_o} \\
                        &=&\alpha_{\rm DR}(p\mhyph\mathrm{H}_3^+) \qquad \qquad \; T \geq 250\,{\rm K}.  \nonumber
\end{eqnarray}\nonumber
\end{subequations}

Table \ref{tab:HHHp} summarizes the values of the principal reaction rate coefficients introduced in the present study, where we have assumed the same branching ratios for reactions\,(\ref{11a}) and (\ref{11b}) as described above.

\begin{table*}[h]
\caption{Principal chemical reactions for $\mathrm{H}_3^+$ formation and destruction.}             
\label{tab:HHHp}      
\begin{center}          
\begin{tabular}{l l l l c c l}     
\hline\hline       
Ref & \multicolumn{3}{c}{Reaction}  &    $A$  $^*$ & $\beta$ & Comment \tabularnewline
   &  \multicolumn{3}{c}{ }  &     (cm$^{3}$ s$^{-1}$) & \tabularnewline
\hline                    
\cite{savic2020}  & $\mathrm{H}_{2}^{+} + \mathrm{H}_{2}$  & $\longrightarrow$ & $\mathrm{H}_{3}^{+} + \mathrm{H}$&  2.27 $\times$ 10$^{-9}$ &  -0.06  & computed from x-section \tabularnewline
 \cite{pagani2009}& $p\mhyph\mathrm{H}_3^+ + e^{-}$ & $\longrightarrow$ &   $\mathrm{H}  + \mathrm{H} + \mathrm{H} $ &  3.5 $\times$ 10$^{-8}$ &  -0.75   &    present fit  \tabularnewline
  \cite{pagani2009} & $p\mhyph\mathrm{H}_3^+ + e^{-}$ &$\longrightarrow$ &   $\mathrm{H}_{2}  + \mathrm{H}$  &  1.75 $\times$ 10$^{-8}$ &  -0.75   & present fit \tabularnewline
  \cite{pagani2009} & $o\mhyph\mathrm{H}_3^+ + e^{-}$ &$\longrightarrow$ &  $\mathrm{H}  + \mathrm{H} + \mathrm{H} $ & 4.0 $\times$ 10$^{-8}$ &   -   & present fit, T $\le$ 250K  \tabularnewline
\cite{pagani2009} & $o\mhyph\mathrm{H}_3^+ + e^{-}$  &$\longrightarrow$ & $\mathrm{H}  + \mathrm{H} + \mathrm{H} $ &  3.5 $\times$ 10$^{-8}$ &  -0.75  &present fit, T $\ge$ 250K \tabularnewline
\cite{pagani2009} &  $o\mhyph\mathrm{H}_3^+ + e^{-}$ &$\longrightarrow$ &  $\mathrm{H}_{2}  + \mathrm{H}$ & 2.0 $\times$ 10$^{-8}$ &   -  &present fit, T $\le$ 250K  \tabularnewline
\cite{pagani2009} & $o\mhyph\mathrm{H}_3^+ + e^{-}$ & $\longrightarrow$ &   $\mathrm{H}_{2}  + \mathrm{H}$  &   1.75 $\times$ 10$^{-8}$ &  -0.75  & present fit, T $\ge$ 250K \tabularnewline
\hline                    
\end{tabular}
\end{center}
$^*$  The reaction rate coefficient, $k$, is expressed as $A \times (T/300) ^{\beta}$. 
\end{table*}
%

\section{Master equation for \texorpdfstring{$\mathrm{H}_3^+$}{H3+} state populations} \label{sec:detailed_balance}

In this section, we describe the master equation for the level populations of $\mathrm{H}_3^+$, and we show that $\mathrm{H}_3^+$ chemistry and specific molecular properties result in an excitation temperature that is lower than the kinetic temperature of the gas.

The demonstration starts from a full treatment of the differential state equations, including all possible processes, i.e., collisional and radiative transitions as well as chemical state-to-state formation and destruction reactions.
Solving these equations, the next section (\ref{Subsec:Theory}) will show that we can indeed recover a value for the $T_{12}(\mathrm{H}_3^+)$ excitation temperature that is systematically below the gas kinetic temperature, similar to observational results. The next sub-section explores how sensitive these equations are to the number of levels included in the computation. This allows to understand why at least $5$ levels must be included for quantitative results. It also shows that the range $[25:50]\,\mathrm{K}$ is much less sensitive to the number of levels included. By restricting our analysis to this temperature range, we can derive a qualitative analytical approximation with only two levels (Section\,\ref{subsec:two_levels}). The resulting expression shows that selective formation of {\em p-$\mathrm{H}_3^+$} by {\em p-$\mathrm{H_2}$} followed by incomplete thermalization is responsible for the deviation from thermal equilibrium.

\subsection{Differential equations} \label{Subsec:Theory}

The variation over time of the density of a level $i$ ($i\in [1,N]$) of $\mathrm{H}_{3}^{+}$, $n_i$, can be described by
\begin{align}
\frac{dn_{i}}{dt} & = k_{\mathrm{form},i} \, n\left(\mathrm{H_{2}^{+}}\right) \, n\left(\mathrm{H}_{2}\right) - \alpha_{\mathrm{DR},i} 
\, n_{i} \, n\left(e^{-}\right)  &  {\rm (formation\,and\,destruction)} \\ \nonumber
 & + \sum_{j\neq i} k_{ji}^{p} \, n\left(\mathrm{H}_{2}\right) \, p_2 \, n_{j} - \sum_{j\neq i} k_{ij}^{p} \, n\left(\mathrm{H}_{2}\right) \, p_2 \, n_{i} & \hfill {\rm (collisions\,with\,} p\mhyph\mathrm{H}_{2}) \\ \nonumber
 & + \sum_{j\neq i}k_{ji}^{o} \, n\left(\mathrm{H}_{2}\right) \, (1-p_2) \, n_{j} - \sum_{j\neq i} k_{ij}^{o} \, n\left(\mathrm{H}_{2}\right) \, (1-p_2) \, n_{i} & \hfill {\rm (collisions\,with\,} o\mhyph\mathrm{H}_{2}) \\ \nonumber
  & + \sum_{j\neq i,X}k_{ji}^{X} \, n\left(\mathrm{X}\right) \, n_{j} - \sum_{j\neq i,X} k_{ij}^{X} \, n\left(\mathrm{X}\right) \, n_{i} & \hfill {\rm (collisions\,with\,other\,species} \, X) \\ \nonumber
 & + \sum_{i<j}A_{ji} \, n_{j} - \sum_{i>j}A_{ij} \, n_{i} & \hfill {\rm (radiative\,transitions)} 
\end{align}

\noindent
with $k_{\mathrm{form},i}$ and $\alpha_{\mathrm{DR},i}$ denoting the state-dependent chemical formation and destruction rates, $k_{ij}^{p,o}$ the collisional excitation/de-excitation rates with $p\mhyph\mathrm{H}_2$ and $o\mhyph\mathrm{H}_2$, and $k_{ij}^{X}$  denote collision rates 
where $X$ stands for $\mathrm{H}$, $\mathrm{He}$, $e^-$. $A_{ij}$ are the radiative emission transition probabilities.

We underline two important points:
\begin{itemize}
   \item The state-dependent chemical formation and destruction rates are introduced in the master equation and have to be evaluated.  It is important to note that the formation rate of a particular level can differ from its destruction rate (see Section~\ref{subsec:Recomb_elec}).
   \item The main formation process of $\mathrm{H}_3^+$ is a highly exothermic reaction, which preferentially populates high energy levels. Thus, chemical formation can be regarded as an excitation mechanism.
\end{itemize}

\noindent
Introducing $x_i$, the relative populations of  $\mathrm{H_{3}^{+}}$,  so that $n_i = x_i \, n\left(\mathrm{H_{3}^{+}}\right)$, the total destruction rate is $\alpha_{\mathrm{DR}} = \sum_i x_i\,\alpha_{\mathrm{DR},i}$. The total formation rate is $k_{\mathrm{form}} = \sum_i k_{\mathrm{form},i}$. Since the relative populations $x_i$ are unknown until the differential equations are solved, it is not possible to compute the destruction rate beforehand,  as long as the $\alpha_{\mathrm{DR},i}$ differ for individual states, and  thus $n\left(\mathrm{H_{3}^{+}}\right)$ cannot be computed independently. At steady state,$\frac{dn_{i}}{dt} = 0 $ and  the system of equations to solve is
\begin{multline}\label{eq:Det_Bal}
\left(\sum_{i>j} A_{ij} + n\left(\mathrm{H}_{2}\right) \, \sum_{j\neq i} \left(k_{ij}^{o} \, (1-p_2) + k_{ij}^{p} \, p_2\right) + \sum_{j\neq i,X} n\left(\mathrm{X}\right) \, k_{ij}^{X} + \alpha_{\mathrm{DR},i} \, n\left(e^{-}\right) \right) \, x_{i} \\
 - \sum_{i<j} A_{ji} \, x_{j} - n\left(\mathrm{H}_{2}\right) \, \sum_{j\neq i} \left(k_{ji}^{o} \, (1-p_2) + k_{ji}^{p} \, p_2 \right) \, x_{j} - \sum_{j\neq i,X} n\left(\mathrm{X}\right) \, k_{ij}^{X} \, x_{j} - k_{\mathrm{form},i} \, R = 0\,,
\end{multline}
with $R=\frac{n\left(\mathrm{H_{2}^{+}}\right)\,n\left(\mathrm{H}_{2}\right)}{n\left(\mathrm{H}_{3}^{+}\right)}$. 
The value of $R$ is initially unknown, as the total destruction rate of  $\mathrm{H}_{3}^{+}$  requires the knowledge of the relative level populations of the molecular ion.
So, we explicitly add  the conservation equation
\begin{equation}
\sum_{i}x_{i} = 1\,.
\label{eq:PopUnity}
\end{equation}
We get a system of $N+1$ equations with $N+1$ unknowns, which is easily solved if the densities of $\mathrm{H_{2}}$, $\mathrm{H_{2}^{+}}$, $e^{-}$ and the temperature are known or assumed. All other quantities are rate coefficients or parameters that are derived from experiment or theory.

\subsubsection*{Derivation of $\mathrm{H_{2}^{+}}$ density and electronic fraction}
It is possible to reduce the set of equations further by estimating $n\left(\mathrm{H_{2}^{+}}\right)$ and $n\left(e^{-}\right)$. The main formation and destruction reactions of $\mathrm{H_{2}^{+}}$ are:
\begin{subequations}
\begin{align}
\mathrm{H}_{2}+\mathrm{CRP} &\longrightarrow \mathrm{H}_{2}^{+}+e^{-}   \hspace*{2.50cm} \zeta \,\,(\mathrm{s}^{-1}),\\
\mathrm{H}_{2}^{+}+\mathrm{H}_{2} &\longrightarrow \mathrm{H}_{3}^{+}+\mathrm{H} \hspace*{2.60cm} k_{\mathrm{form}} \,\,(\mathrm{cm}^3\,\mathrm{s}^{-1}), \\
\mathrm{H}_{2}^{+}+e^{-} &\longrightarrow \mathrm{H}+\mathrm{H}   \hspace*{2.90cm} \alpha^{\mathrm{H_{2}^{+}}} \,\,(\mathrm{cm}^3\,\mathrm{s}^{-1}),
\end{align}
\end{subequations}
 where $\mathrm{CRP}$ represents cosmic ray particles and $\zeta$ the corresponding ionization rate of $\mathrm{H}_{2}$. At steady state, this leads to (with $\alpha^{\mathrm{H_{2}^{+}}}$ the dissociative recombination rate of $\mathrm{H_{2}^{+}}$)
\begin{equation}
n\left(\mathrm{H_{2}^{+}}\right) = \frac{\zeta \, n\left(\mathrm{H}_{2}\right)}{k_{\mathrm{form}} \, n\left(\mathrm{H}_{2}\right) + \alpha^{\mathrm{H_{2}^{+}}} \, n\left(e^{-}\right)}.
\end{equation}
This removes one parameter from the system.

As for $n\left(e^{-}\right)$, in diffuse gas it is often assumed to be equal to the density of $\mathrm{C}^+$. However, for high cosmic ray ionization rates, as, e.g., in the Central Molecular Zone (CMZ) of our galaxy, Le Petit et al. \cite{lepetit2016} have shown that protons may contribute significantly to the ionization fraction. Here we compute the electronic density considering both $\mathrm{H}^+$ and $\mathrm{He}^+$ as described in App.\,\ref{ab:xe},
which involves the relative abundances of all atoms with ionization potential below the Lyman cutoff (assumed to be fixed) and the UV radiation field \footnote{The UV radiation field strength $G_0$ is defined in \cite{draine1978}. It controls the interstellar grain charge, which impacts the recombination of $\mathrm{H}^+$ and $\mathrm{He}^+$.}.
With these derivations of $n\left(\mathrm{H_{2}^{+}}\right)$ and $n\left(e^{-}\right)$, the system formed by Eq.\,(\ref{eq:Det_Bal}) and Eq.\,(\ref{eq:PopUnity}) depends on five astrophysical parameters: $n_{\mathrm{H}}$\footnote{$n_{\mathrm{H}}$, the proton density is expressed as = ${n(\mathrm{H}) + 2\,n(\mathrm{H}_2) + n(\mathrm{H}^+)}$.}, $T$, $f_{m}$, $G_{0}$, 
and $\zeta$, where we introduced $f_{m}$ the molecular fraction, $f_{m} = 2 \, n(\mathrm{H}_2) / n_{\mathrm{H}}$. 

\subsection{Numerical solution of the coupled equations}
We developed \verb=ExcitH3p=, a FORTRAN code that solves the $\mathrm{H}_3^+$ coupled system of equations (Eq.\,\ref{eq:Det_Bal}) for any number of levels, and then computes the two-level excitation temperature (Eq.\ref{T01H3p})
\begin{equation}
T_{12}\left(\mathrm{H}_3^+\right) =    32.86 \,\mathrm{K} / \ln\left(\frac{2\, x_{1}}  { x_{2}}\right).
\end{equation}
Here $x_1$ and $x_2$ denote the relative populations of the $(1,1)$ ground state and the first excited state $(1,0)$, respectively. 
\begin{figure}
\begin{center}
\includegraphics[width=0.95\textwidth]{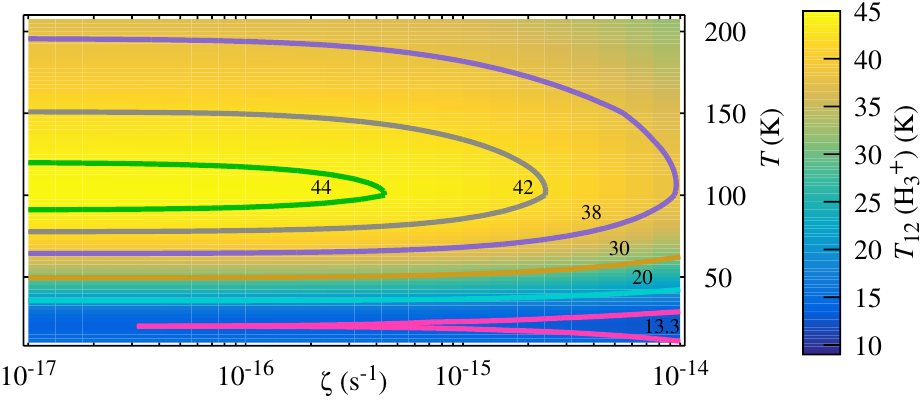}
\end{center}
\caption{The excitation temperature $T_{12}(\mathrm{H}_3^+)$ as calculated using \texttt{ExcitH3p}, as a function of $\zeta$ and $T$ for a density $n_\mathrm{H} = 10^2\,\mathrm{cm}^{-3}$ and a molecular fraction $f_m = 0.8$. Note that the lowest 24 levels of ${\mathrm{H}_3^+}$ are included in the model, while only the lowest two states are used to determine $T_{12}(\mathrm{H}_3^+)$. \label{T_exc_H3p}}
\end{figure}

In practice, the $24$ energetically lowest rotational levels of $\mathrm{H}_3^+$ are included in the model. Those are the levels for which radiative emission rates as well as collisional excitation/de-excitation rates are known or have been estimated. The highest level in this framework is the $(7,6)$  {\em ortho} level, located $2190\,\mathrm{K}$ above the ground state.
We find that the solution to the set of equations (\ref{eq:Det_Bal}) recovers very nicely the full results obtained with the Meudon PDR code for diffuse line of sight conditions. The advantage of the master equation approach is that it is much less computationally expensive, and allows for a rapid exploration of the parameter space and the various possible hypotheses concerning the less well-known physical processes.

Results derived by the FORTRAN routine are presented in Figure\,\ref{T_exc_H3p} for a range of typical values of the cosmic ionization rate $\zeta$ and gas kinetic temperature $T$. The total gas density is set to $n_\mathrm{H} = 10^2\,\mathrm{cm}^{-2}$ and the molecular fraction to $f_m = 0.8$. We find that in the entire parameter space, $T_{12}(\mathrm{H}_3^+)$, the excitation temperature given by the two lowest $\mathrm{H}_3^+$ levels, is systematically lower than the kinetic temperature. The overall magnitude of the discrepancy between the two temperatures is in agreement with the observations \cite{indriolo2012,albertsson2014}. This outcome is a sole property of the microscopic excitation and de-excitation mechanisms of the individual quantum levels
of $\mathrm{H}_3^+$, which are detailed above. 
 Figure\,\ref{T_exc_H3p} also shows that 
 a maximum of about $44\,\mathrm{K}$ 
 is reached for $T_{12}(\mathrm{H}_3^+)$ at kinetic temperatures around  $100\,\mathrm{K}$. The occurrence of such a maximum is remarkable. We have verified that the maximum is still there, although somewhat different, when the dissociative recombination rate coefficients of {\em para} and {\em ortho}-$\mathrm{H}_3^+$ are the same:  $T_{12}(\mathrm{H}_3^+)^{max}$ = 38 K for T$_{gas}$ = 150 K under the same physical conditions. We come back to that point in Section 4.2.1 when discussing the influence of the number of ${\mathrm{H}_3^+}$  levels included in the model.

\begin{figure}
\begin{center}
\includegraphics[width=0.95\textwidth]{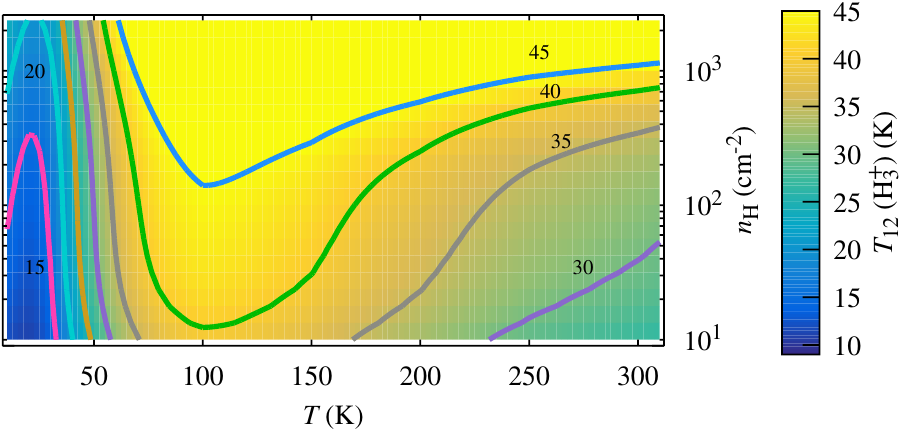}
\end{center}
\caption{The excitation temperature $T_{12}(\mathrm{H}_3^+)$ as a function of the gas kinetic temperature $T$ and density $n_{\mathrm{H}}$ for $\zeta = 10^{-16}\,\mathrm{s}^{-1}$, a molecular fraction $f_m = 0.8$
and considering the first 24 levels of ${\mathrm{H}_3^+}$. \label{T_exc_H3p_dens}}
\end{figure}

 Figure\,\ref{T_exc_H3p_dens} shows the influence of density, showing that $T_{12}\left(\mathrm{H}_3^+\right)$ is not sensitive to this parameter up to densities of more than $10^3\,\mathrm{cm}^{-3}$ if the gas temperature is below $\sim$ 80K.

\subsubsection{Influence of the number of ${\mathrm{H}_3^+}$ levels included in the model} 
\begin{figure}
\begin{center}
\includegraphics[width=0.95\textwidth]{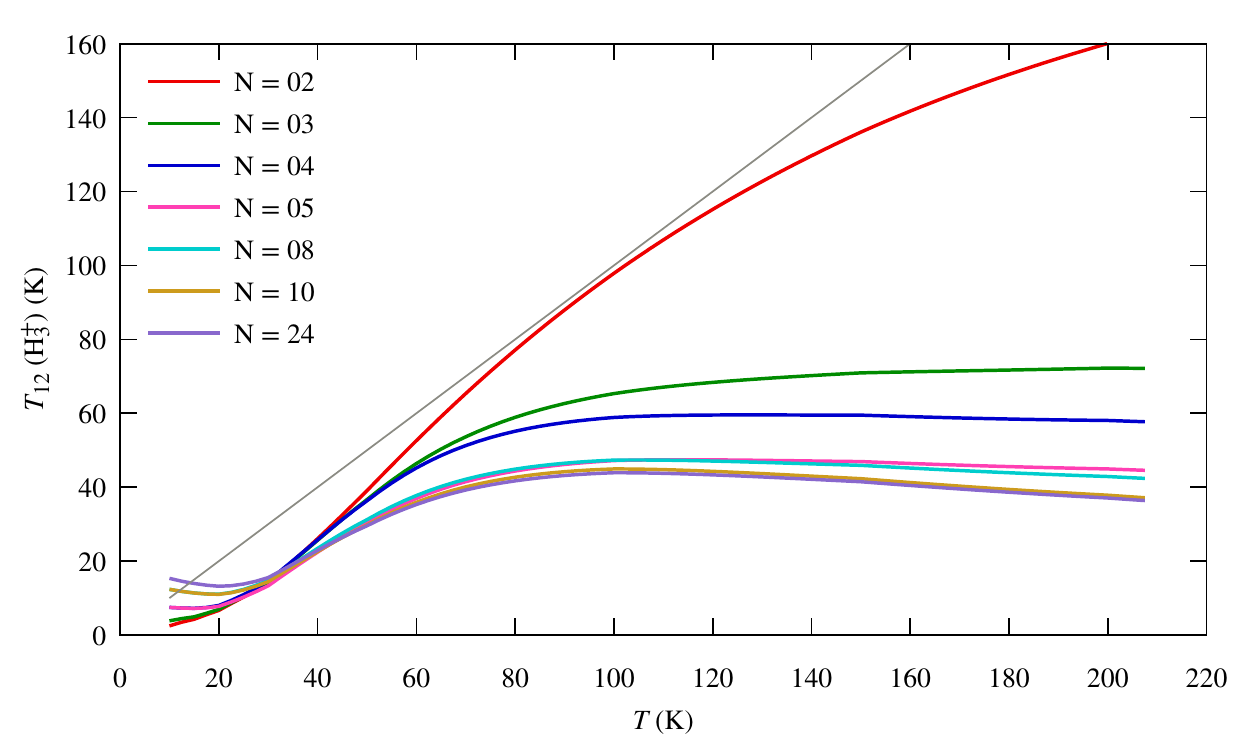}
\end{center}
\caption{$T_{12}(\mathrm{H}_3^+)$ as a function of the number of levels taken into account in the coupled set of equations\,(\ref{eq:Det_Bal}). We assume values of $n_\mathrm{H} = 10^2\,\mathrm{cm}^{-3}$, $\zeta = 5\,10^{-16}\,\mathrm{s}^{-1}$, $f_m = 0.8$. The straight line in gray is the first diagonal showing complete thermalization. \label{Fig:N_levels}}
\end{figure}

Interesting conclusions can be drawn from the examination of the dependence of our results on the number $N$ of ${\mathrm{H}_3^+}$ levels included in the model. Except for the Galactic center, only the two lowest rotational levels of H$_3^+$ have been detected in the ISM so far. 
Consequently, the interpretation of astrophysical observations is usually restricted to these two levels. Figure\,\ref{Fig:N_levels} displays the computed $T_{12}(\mathrm{H}_3^+)$ value as a function of the kinetic temperature $T$, for different values of $N$.
We perform this comparison for the following parameters: $\zeta=5\,10^{-16}\,\mathrm{s}^{-1}$, $n_\mathrm{H} = 100\,\mathrm{cm}^{-3}$, and $f_m = 0.8$. 
We find that the  curve corresponding to $N = 2$
is far from the values obtained with $24$ levels, except for a small range at very low temperatures. Nevertheless, even with $N = 2$, $T_{12}(\mathrm{H}_3^+)$ is systematically below the kinetic temperature.
$N = 5$ is the minimum needed for acceptable results, and convergence is reached for $N = 10$ for the physical conditions considered here. It is important to realise that $N = 5$ involves the metastable $(3,3)$ level, representing a possible sink for $\mathrm{H}_3^+$ population.

The decrease of $T_{12}(\mathrm{H}_3^+)$ with increasing gas kinetic temperature above $100\,\mathrm{K}$ seems puzzling at first. However, it can be understood by considering the various possible decay paths from the two excited {\em para} levels $(2,2)$ (corresponding to relative population $x_3$ in our enumeration) and $(2,1)$ (corresponding to $x_4$), and the lowest excited {\em ortho} level $(3,3)$ (corresponding to $x_5$).
It is important to note that the $(3,3)$ level is metastable with  an infinite radiative lifetime. 
Hence it can only be depopulated in collisional processes. Examination of the collision rates with $\mathrm{H}_2$ shows that its branching ratios towards {\em ortho} and {\em para} are approximately equal in the $30 - 300\,\mathrm{K}$ temperature range. Both {\em para} levels $(2,2)$ and $(2,1)$, on the other hand, can decay both radiatively and collisionally. The critical densities of these levels, given by the ratio of the radiative emission rate and the total collisional de-excitation rate coefficient, are a few thousand cubic centimeters. 
Thus, at densities of a few hundred cubic centimeters, relevant for the astrophysical observations discussed here, radiative decay of {\em para} levels to the ground state is much more likely than collisional de-excitation, while the ground {\em ortho} level (1,0) is underpopulated compared to a thermal Boltzmann distribution, as population may be trapped in the $(3,3)$ level. 
This leads to a perceived overpopulation of the lowest {\em para} state, compared to the lowest {\em ortho} state. Chemical formation can populate efficiently all levels due to the large exothermicity of the reaction. So, increasing $N$ leads to more open channels to populate levels that decay to $(3,3)$ and inhibits population of the lowest {\em ortho} state $(1,0)$.

\subsection{Two-level approximation} \label{subsec:two_levels}

In a restricted range of kinetic temperatures around $T \simeq 25-50\,\mathrm{K}$, the two-level case is a fair approximation to the full system, as seen in Figure\,\ref{Fig:N_levels}. It allows to considerably simplify the system and to derive an analytic expression for $x_2/x_1$ that offers the opportunity to highlight the key microscopic mechanisms at work. 

We restrict our study to molecular hydrogen collisions and introduce $k_{12} = k_{12}^{o} \, (1-p_2) + k_{12}^{p} \, p_2$, the total collisional excitation  rate due to $\mathrm{H}_2$ from level $1$ to level $2$ (and the equivalent for $k_{21}$).
Then, the coupled equations system (Eq.\,\ref{eq:Det_Bal}) reduces to:
\begin{subequations}
\begin{align}
\left(n\left(\mathrm{H}_{2}\right) \, k_{12} + \alpha_{\mathrm{DR},1} \, n\left(e^{-}\right)\right) \, x_{1} - n\left(\mathrm{H}_{2}\right) \, k_{21} \, x_{2} &= k_{\mathrm{form},1} \, R \\
- n\left(\mathrm{H}_{2}\right) \, k_{12} \, x_{1} + \left(n\left(\mathrm{H}_{2}\right) \, k_{21} + \alpha_{\mathrm{DR},2} \, n\left(e^{-}\right)\right) \, x_{2} &= k_{\mathrm{form},2} \, R
\end{align}
\end{subequations}

Introducing $k_{\mathrm{form}} =  k_{\mathrm{form},1} + k_{\mathrm{form},2} $, we can compute the $x_{2}/x_{1}$ ratio: 
\begin{equation}
\label{Eq:x2sx1}
\frac{x_{2}}{x_{1}} = \frac{n\left(\mathrm{H}_{2}\right) \, k_{12} \, k_{\mathrm{form}} + \alpha_{\mathrm{DR},1} \, n\left(e^{-}\right) \, k_{\mathrm{form},2}}{n\left(\mathrm{H}_{2}\right) \, k_{21} \, k_{\mathrm{form}} + \alpha_{\mathrm{DR},2} \, n\left(e^{-}\right) \, k_{\mathrm{form},1}} \, ,
\end{equation}

The factor $R$ cancels out and  the ratio $x_{2} / x_{1}$ does not depend on $n\left(\mathrm{H}_{3}^+\right)$. We introduce the electronic fraction $x_e = n\left(\mathrm{e}^-\right)/n_\mathrm{H}$ and the configuration-specific formation rates of H$_3^+$ with $k_{\mathrm{form},1} = p_3^f \,k_{\mathrm{form}}$  and $k_{\mathrm{form},2} = (1-p_3^f) \,k_{\mathrm{form}}$. Furthermore, we apply detailed balance to the H$_3^+$--H$_2$ collisional rates, yielding $k_{12}=\frac{g_{2}}{g_{1}}\,\exp\left(-\frac{E_{21}}{T}\right)\,k_{21}$.
Finally, we get the following expression
\begin{equation}
\frac{x_{2}}{x_{1}} = \frac{g_{2}}{g_{1}} \, \exp\left(-\frac{E_{21}}{T}\right) \, \left[ \frac{1 + \left(1 - p_2\right) \, \frac{4}{3} \, \frac{\alpha_{\mathrm{DR},1}}{k_{21}} \, \frac{x_e}{f_m} \, \frac{g_{1}}{g_{2}} \, \exp\left(\frac{E_{21}}{T}\right)}{1 + \left(1 + 2\,p_2\right) \, \frac{2}{3} \, \frac{\alpha_{\mathrm{DR},2}} {k_{21}} \, \frac{x_e}{f_m}}\right]
=\frac{g_{2}}{g_{1}} \, \exp\left(-\frac{E_{21}}{T_{12}\left(\mathrm{H}_3^+\right)}\right)\,.
\label{x1sx2}
\end{equation}
Apart from the kinetic temperature $T$, the $x_2/x_1$ ratio depends on
the collisional and dissociative recombination  rate coefficients, the molecular fraction $f_m$ of $\mathrm{H}_{2}$, and the electronic fraction $x_e$.
It is independent of the cosmic ray ionization rate $\zeta$ and of the {\em total} formation reaction rate coefficient of H$_3^+$, but -- crucially -- not of the branching ratios, which effectively enter the equation through the H$_2$ {\em para} fraction $p_2$.

We can interpret the term in square brackets as a correction factor to the Boltzmann value for the kinetic temperature. At the low temperatures considered here $(25-50\,\mathrm{K})$ the value of $p_2$ ranges from $0.6$ to $1$ (see Figure\,\ref{fig2}), and numerical analysis reveals that the $(1-p_2)$ and $(1+2\,p_2)$ pre-factors in the nominator and the denominator are sufficient to keep the overall correction factor smaller than unity in the entire temperature range, resulting in a lowering of the $\frac{x_2}{x_1}$ ratio compared to the thermal value (in agreement with astronomical observations). The extent of the deviation from thermal equilibrium depends critically on the ratios $\frac{\alpha_{\mathrm{DR},1}}{k_{21}}$ and $\frac{\alpha_{\mathrm{DR},2}}{k_{21}}$ of the electron recombination rates to the collisional rate coefficients. If we, for the sake of argument, extrapolate the formula using an artificially large value for $k_{21}$ -- corresponding to very efficient collisional thermalization -- the entire correction factor tends toward unity, and we will recover the ratios given by the gas kinetic temperature. The same is obviously true for very small DR rate coefficients $\alpha_{\mathrm{DR},1}$ and $\alpha_{\mathrm{DR},2}$. 

The picture that emerges is that the excitation temperature of $\mathrm{H}_3^+$ appears lower than the nominal gas temperature because of incomplete thermalization. The formation process strongly favors the formation of $p\mhyph\mathrm{H}_3^+$ at low temperatures, as the {\em para}-fraction $p_2$ of $\mathrm{H}_2$ is large. The electron recombination process then removes H$_3^+$ before it can reach thermal equilibrium in collisions with $\mathrm{H}_2$. While we stress that these conclusions based on the two-level approximation are only valid in a limited temperature range, and that for quantitative results more levels need to be considered, we can reproduce the general trend very well using the master equation approach described above. For artificially enlarged $\mathrm{H}_3^+ - \mathrm{H}_2$ collisional rate coefficients (or sufficiently reduced electron recombination rates) the calculations reach complete thermal equilibrium between the excitation temperature $T_{12}\left(\mathrm{H}_3^+\right)$ and the kinetic temperature $T$.      

In essence, the nuclear spin restrictions of the $\mathrm{H}_3^+$ formation reaction produce an over-proportional amount of $\mathrm{H}_3^+$ in the {\em para} configuration, and thermalization in collisions with H$_2$ is too slow to reach equilibrium before the ions are destroyed by free electrons. This is to be contrasted to the situation for $\mathrm{H}_2$, where the destruction process is much slower compared to thermalizing collisions (see Appendix\,\ref{App:TwoLevelH2} for details).  

\section{Comparison to observations} \label{sec:results}

\subsection{Master equation approach} \label{0D_model}

To validate our approach, we compare the results of the master equation approach (Eq.\,\ref{eq:Det_Bal}) with observations. Table\,\ref{tab:Observations} presents the $\mathrm{H}_3^+$ excitation temperature reported in the literature for a few local diffuse clouds as well as the corresponding $\mathrm{H}_2$ excitation temperatures. Data for $\mathrm{H}_2$ come from Copernicus \cite{savage1977} and FUSE \cite{rachford2002,rachford2009} satellite  observations. The proton densities reported in this table are those published in the papers reporting the $\mathrm{H}_2$ data. Determination of diffuse cloud density with observations of $\mathrm{H}$ and $\mathrm{H}_2$ is not straightforward, because a fraction of the hydrogen atoms observed on the line of sight may not be related to the diffuse clouds to which $\mathrm{H}_2$ belongs. So, these densities are to be seen as order of magnitude estimates, and should be considered only as indicative.  

\begin{table*}[h]
\caption{Selection of sightlines with both H$_2$ and H$_3^+$ observations. }
\label{tab:Observations}
\begin{center}
\begin{tabular}{llcccccl}
\hline
\hline
         &                 & $n_{\mathrm{H}}$                & $T_{01}^{obs}\left(\mathrm{H}_{2}\right)$ & $T_{12}^{obs}\left(\mathrm{H}_{3}^{+}\right)$ & $N\left(\mathrm{H}_{3}^{+}\right)$                & $N\left(\mathrm{H}_{2}\right)$                   & References\tabularnewline                               
         &                 & $\left(\mathrm{cm}^{-3}\right)$ & $\left(\mathrm{K}\right)$           & $\left(\mathrm{K}\right)$                & $\left(10^{13}\thinspace\mathrm{cm}^{-2}\right)$  & $\left(10^{20}\thinspace\mathrm{cm}^{-2}\right)$ &           \tabularnewline                               
HD154368 &                 & $240$                                           & $51 \pm 8$                                                  & $20 \pm 4$                                                        & $9.37$                                                                              & $14.4$ & \cite{crabtree2011,indriolo2012,rachford2002} \tabularnewline     
HD73882  &                 & $520$                                           & $51 \pm 6$                                                  & $23 \pm 3$                                                        & $9.02$                                                                              & $12.9$ & \cite{crabtree2011,indriolo2012,rachford2002} \tabularnewline
HD27778  & 62 Tau          & $280$                                           & $55 \pm 7$                                                  & $29 \pm 4$                                                        & $6.49$                                                                              & $6.23$ & \cite{albertsson2014,rachford2002} \tabularnewline
HD24398  & $\zeta$ Per     & $215$                                           & $57 \pm 6$                                                  & $28 \pm 4$                                                        & $6.26$                                                                              & $4.75$ & \cite{crabtree2011,indriolo2012,savage1977} \tabularnewline
HD24534  & X Per           & $325$                                           & $57 \pm 4$                                                  & $46^{+21}_{-13}$                                                   & $7.34$                                                                              & $8.38$ & \cite{crabtree2011,indriolo2012,rachford2002} \tabularnewline
HD41117  & $\chi^{2}$ Ori  & $200$                                           & $60 \pm 7$                                                  & $29 \pm 13$                                                       & $5.29$                                                                              & $4.90$ & \cite{albertsson2014,rachford2009} \tabularnewline
HD110432 &                 & $140$                                           & $68 \pm 5$                                                  & $30 \pm 2$                                                        & $5.22$                                                                              & $4.37$ & \cite{crabtree2011,indriolo2012} \tabularnewline
HD210839 & $\lambda$ Cep   & $115$                                           & $72 \pm 6$                                                  & $34 \pm 2$                                                        & $7.58$                                                                              & $6.88$ & \cite{indriolo2012,rachford2002} \tabularnewline
HD43384  & 9 Gem           & \emph{120}$^{*}$                                & $74 \pm 15$                                                 & $38 \pm 11$                                                       & $4.07$                                                                              & $7.36$ & \cite{albertsson2014,rachford2009} \tabularnewline
\hline
\end{tabular}
\par\end{center}
$^{*}$ value unavailable. Educated guess only.
\end{table*}

\begin{figure}
\begin{center}
\includegraphics[width=0.95\textwidth]{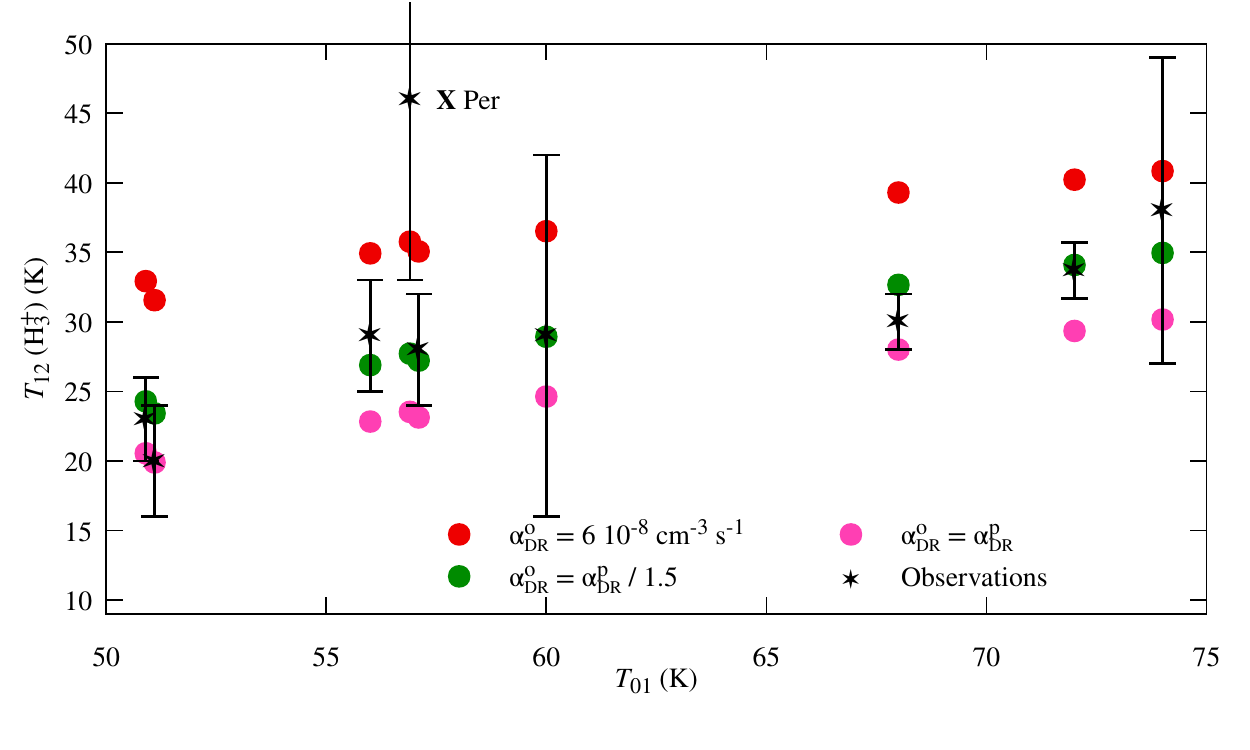}
\end{center}
\caption{Computed H$_3^+$ excitation temperature $T_{12}(\mathrm{H}_3^+)$ as a function of the observed $\mathrm{H}_2$ excitation temperature for the $9$ lines of sight given in Tab.\,\ref{tab:Observations}. Sightlines with the same $T_{01}$ have been shifted by $0.1\,\mathrm{K}$ for clarity. Three different hypothesis are used for the dissociative recombination rate of $o\mhyph\mathrm{H}_3^+$ with electrons (see text). \label{T_exc_Model}}
\end{figure}

Using the \verb=ExcitH3p= program that solves Eq.\,(\ref{eq:Det_Bal}), we compute $T_{12}\left(\mathrm{H}_3^+\right)$ for all the lines of sight. We assume a cosmic ray ionization rate of $\zeta = 10^{-16}\,\mathrm{s}^{-1}$ and a molecular fraction of $f_m = 0.8$. For the gas density, the values in Tab. \ref{tab:Observations} are used, and for the gas temperature we  use the values derived from $\mathrm{H}_2$ observations $T_{01}^{obs}\left(\mathrm{H}_2\right)$.
\subsubsection*{Model results and sensitivity to the DR rate coefficients}
Figure\,\ref{T_exc_Model} shows the results for three different hypothesis for $\alpha_{DR}(o\mhyph\mathrm{H}_3^+)$, and the value provided by equation\,(\ref{alp_DR_p}) for $\alpha_{DR}(p\mhyph\mathrm{H}_3^+)$. We see a quasi-linear variation of $T_{12}\left(\mathrm{H}_3^+\right)$ with $T_{01}$, which is well reproduced by the models.
Using $\alpha_{DR}$ from equation.\,(\ref{alp_DR_o}) (red circles) leads to temperatures which are too high, while using the same rate for $o\mhyph\mathrm{H}_3^+$ and $p\mhyph\mathrm{H}_3^+$ (pink circles) leads to temperatures which are too low. An empirical adjustment using $\alpha_{\rm DR}(p\mhyph\mathrm{H}_3^+)$ from Eqs.\,(\ref{alp_DR_p}) and $\alpha_{\rm DR}(o\mhyph\mathrm{H}_3^+) = \alpha_{\rm DR}(p\mhyph\mathrm{H}_3^+) / 1.5$ (green circles) gives a very satisfying result, given that no attempt has been made to optimize other parameters.

There is a single 
noticeable exception: $\mathrm{X}$ Per. However, $T_{12}\left(\mathrm{H}_3^+\right)$ of this line of sight suffers from a particularly large error bar as listed in Table \ref{tab:Observations} . 
A high excitation temperature of  $\mathrm{H}_3^+$ can only be reached for a kinetic temperatures close to $100\,\mathrm{K}$ with our models, as can be seen in Figure\,\ref{T_exc_H3p}.

We note that typically $10$ to $12\,\%$ of $\mathrm{H}_3^+$ ions are in excited stable or metastable levels above the respective lowest \emph{ortho} or \emph{para} levels, such as $(3,3)$, $(4,4)$, $(5,5)$, $(6,6)$ and $(7,7)$.

\subsection{Full PDR model} \label{PDR_Model}

For comparison and validation, we compare the results of Section\,\ref{0D_model} to those obtained with our full PDR code for the case of $\zeta$ Per. We do not seek a complete model of this line of sight, and do not try to optimize the free parameters to reproduce other observations than $\mathrm{H}_2$ and $\mathrm{H}_3^+$ excitation.

We use our previous study dedicated to $\zeta$ Per \cite{lepetit2004} as a starting point, but  we also account for the many updates made since then (\citep{lepetit2006,lebourlot2012,lepetit2016}).
For the present study, we have introduced in the Meudon PDR code the {\em ortho/para} dependence of the formation reaction of $\mathrm{H_3^+}$ via $\mathrm{H_2}$ + $\mathrm{H_2^+}$ collisions and the nuclear spin dependence of the dissociative recombination rate
coefficient, in addition to the other excitation mechanisms of  $\mathrm{H_3^+}$. The collisional excitation of $\mathrm{H_2}$ by $\mathrm{H^+}$, that has been revisited by \cite{gonzales2021} for highly rovibrationally excited levels, has also been updated.
Observations suggest a total visual extinction $A_\mathrm{V} = 0.9\,\mathrm{mag}$ ($R_\mathrm{V} = 2.8$, $E_{B-V} = 0.32$, $N_\mathrm{H}/E_{B-V} = 5.2\,10^{21}\,\mathrm{cm}^{-2}$). In our models the cloud is illuminated from both sides by the standard ISRF ($G_0 = 1$).

We consider two different scenarios
\begin{description}
    \item [Model A] Constant density and constant temperature,
    \item [Model B] Constant density, with computation of the thermal balance.
\end{description}

Both models A and B are fairly standard. 
To account for the presence of purely atomic gas along the line of sight, we use an $A_\mathrm{V} = 0.7$ for both models A and B, allowing to account for the molecular hydrogen abundance.

The results are summarised in Table\,\ref{PDR_code}. The examples shown here have been selected from an evaluation of the following $\chi^2$, where $\sigma$ are the observational uncertainties
\begin{multline}
\chi^2 = \frac{1}{4}\, \left( 
\frac{(N_{Obs}(\mathrm{H}_2) - N_{Mod}(\mathrm{H}_2))^2}{\sigma_{N(\mathrm{H}_2)}^2} +
\frac{(N_{Obs}(\mathrm{H}_3^+) - N_{Mod}(\mathrm{H}_3^+))^2}{\sigma_{N(\mathrm{H}_3^+)}^2} + \right. \\
\left. \frac{(T_{01,Obs}(\mathrm{H}_2) - T_{01,Mod}(\mathrm{H}_2))^2}{\sigma_{T_{01}(\mathrm{H}_2)}^2} +
\frac{(T_{12,Obs}(\mathrm{H}_3^+) - T_{12,Mod}(\mathrm{H}_3^+))^2}{\sigma_{T_{12}(\mathrm{H}_3^+)}^2}
\right).
\end{multline}
%

\begin{table*}[h]
\caption{PDR model results. Column densities are in $\mathrm{cm}^{-2}$ and excitation temperatures in $\mathrm{K}$. Numbers in parenthesis are power of $10$.}
\label{PDR_code}
\begin{center}
\begin{tabular}{cccccc}
\hline
\hline
Obs. & $N(\mathrm{H}_2)$  & $T_{01}$ & $N(\mathrm{H}_3^+)$ & $T_{12}(\mathrm{H}_3^+)$ & $\chi^2$ \tabularnewline
\hline
$\zeta$ Per  & 4.75(20)  & $57 \pm 6$ &$6.26(13)$ & 28 $\pm$ 4   & \tabularnewline
 \cite{indriolo2012} & $\pm0.95(20)$ & &$\pm0.52(13)$  &  & \tabularnewline
\hline 
 Models &  \multicolumn{4}{c}{ $n_\mathrm{H} = 150\,\mathrm{cm}^{-3}$, $\zeta = 8\,10^{-16}\,\mathrm{s}^{-1}$} & \tabularnewline
\hline

 A & $5.00(20)$ & $56.9$ & $5.62(13)$ & $31.0$ & $0.54$ \tabularnewline
 B & $5.00(20)$ & $62.3$ & $6.30(13)$ & $33.7$ & $0.73$ \tabularnewline
\hline
   &  \multicolumn{4}{c}{ Same parameters, but $\alpha_{DR}^o = \alpha_{DR}^p / 1.5$} & \tabularnewline
\hline
 A & $5.00(20)$ & $56.9$ & $4.58(13)$ & $23.0$ & $3.0$ \tabularnewline
 B & $5.00(20)$ & $62.3$ & $5.18(13)$ & $26.0$ & $1.4$ \tabularnewline
\hline
\end{tabular}
\par\end{center}
\end{table*}

The proposed ionization rates are rather high, but in line with the other recent evaluations based on $\mathrm{H}_3^+$ and $\mathrm{OH}^+$abundances \citep{indriolo2010,indriolo2012,bacalla19}.
We note that our previous estimate of the cosmic ionization rate
towards $\zeta$ Per was somewhat lower, as we included in that study additional constraints provided by $\mathrm{OH}$ and $\mathrm{HD}$ column densities. The constraints imposed by the excitation temperature of  $\mathrm{H}_2$ and $\mathrm{H}_3^+$ 
are not sensitive to the cosmic ionization rate of $\mathrm{H}_2$, as shown in Fig. \ref{T_exc_H3p}.
We then recover the predictions based on molecular ion observations.

Using the recombination rate from Eq.\,(\ref{alp_DR_o}), the excitation temperature of $\mathrm{H}_3^+$ is slightly over-estimated, as found in the previous Section. Applying the same empirical correction to the recombination rate leads to a lower excitation temperature, without any impact on $\mathrm{H}_2$. The total amount of $\mathrm{H}_3^+$ is lower due to the overall larger recombination rate. The resulting $\chi^2$ varies accordingly. This illustrates the very high sensitivity of $\mathrm{H}_3^+$ to the exact value of the electron recombination rate. The situation now is much better than it used to be, but smaller uncertainties are still needed for quantitative analysis. Note that we do not claim to estimate these rates from observational data.

Overall, this comparison validates the excitation model presented in Section\,\ref{sec:detailed_balance} and shows that the temperatures of $\mathrm{H}_3^+$ and $\mathrm{H}_2$ in the diffuse ISM can, in fact, be predicted considering a a strongly reduced set of reactions.

\section{Discussion and conclusion} \label{sec:discussion}

In this paper, we review all physical processes relevant to formation, excitation and destruction of $\mathrm{H}_3^+$ in the diffuse interstellar medium. We provide references to the best data available to date.

We show that a $0D$ statistical model of $\mathrm{H}_3^+$ level populations, including formation/destruction terms and updated collisional excitation/deexcitation processes, allows to explain the low excitation temperature observed in diffuse clouds that has puzzled observers and modelers alike for twenty years. In particular, we show that it is mandatory to include state-to-state chemical formation and destruction processes to explain the departure from thermal equilibrium (Boltzmann ratio) observed for the two lowest levels of $\mathrm{H}_3^+$. Specifically, the formation of $p\mhyph\mathrm{H}_3^+$ by $p\mhyph\mathrm{H}_2$ at temperatures below $70\,\mathrm{K}$ is efficient. Considering the individual levels of $\mathrm{H}_3^+$, we find that reactive collisions with $\mathrm{H}_2$ are generally too slow -- when compared to spontaneous radiative decay and the fast destruction by electron recombination -- to bring the populations into equilibrium with the gas kinetic temperature. These results are confirmed by an updated version of the Meudon PDR code that includes the specific {\em ortho}/{\em para}-dependence of the formation/destruction reactions of $\mathrm{H}_3^+$ in addition to the radiative/collisional excitation balance of that molecular ion. 

While the formation process may be primarily responsible for the increased population of \mbox{$p$-H$_3^+$} levels at low temperature, it is important to note that the consideration of different classes of processes -- radiative transitions as well as chemical reactions with H$_2$ -- is required to achieve quantitative results. We find that all attempts to simplify our master equation further and remove more processes from our models lead to significant changes in the H$_3^+$ excitation temperature and impair the agreement with the observational data. 

Moreover, we show that the inclusion of rotationally excited levels, besides the respective {\em ortho} and {\em para} ground states that are usually considered, has substantial implications for the population of the first two levels and thus for $T_{12}\left({\rm H}_3^+\right)$, and we find that at least 10 levels should be included in the coupled equations to get a converged result.
Our models suggest that typically more than $10\,\%$ of $\mathrm{H}_3^+$ ions are in metastable excited states, which may be observable in absorption towards bright stars or quasi-stellar objects.

Finally, we stress that some key processes are still either badly determined or completely unknown. In particular, precise quantitative computation of $\mathrm{H}_3^+$ excitation will not be possible as long as we still lack accurate state specific recombination rates with electrons at temperatures between $20$ and $200\,\mathrm{K}$. Numerical manipulations of the rate coefficients shown in Section\,\ref{sec:results} reveal the sensitivity of $T\left( \mathrm{H}_3^+\right)$ to the ratio $\alpha_{DR}^o / \alpha_{DR}^p$. However, it is well-known how dangerous it can be to try to infer reaction rates from observational results, and we do not claim that the rates used here are the final word. Another class of processes that are lacking accurate description are collision rates of $\mathrm{H}_3^+$ with $\mathrm{He}$ and $\mathrm{H}$. In particular, exchange of hydrogen atoms during collisions with $\mathrm{H}$ may impact the {\em ortho} to {\em para} ratio of $\mathrm{H}_3^+$. Here we used estimated values for the rate coefficients scaled from reaction rates with $p\mhyph\mathrm{H}_2$ in order to include these processes in the models. While our results seem not to depend strongly on the exact choice of the estimated rate coefficients, more accurate values for these reactions are  clearly desirable.

Despite the remaining limitations, our models for the first time are able to account for the observed $\mathrm{H}_3^+$ excitation temperature in diffuse cloud sightlines. This marks a major step in our understanding of interstellar hydrogen chemistry, providing a framework of state-selective chemistry for two of the most important and fundamental  molecular gas phase species. 

\pagebreak
\appendix

\section{Two-level  approximation for \texorpdfstring{$\mathrm{H}_2$}{H2}}
\label{App:TwoLevelH2}

As first recognized by (\cite{dalgarno1973,spitzer1973}) from
Copernicus observations, $T_{01}$ of $\mathrm{H}_2$ is an excellent proxy for the kinetic temperature $T$ in diffuse clouds, where
collisions with protons allow the two rotational levels to reach thermal equilibrium, in absence of any radiative transition. 
This is not true anymore in dense cloud conditions, where the {\em ortho}-to-{\em para} ratio is expected to be very far from thermal equilibrium \cite{faure2019}, and where collisions with $\mathrm{H}_3^+$ 
modify the excitation balance.
We discuss rapidly the conditions of validity of this feature through a simple 2-level approximation. 
Transitions between $J=0$ and $J=1$ occur only through reactive collisions with $\mathrm{H}^+$ with rates $k_{01}$ and $k_{10}$ (reactive collisions with other species ($\mathrm{H}$, $\mathrm{H}_3^+$) are negligible here). Besides collisions, these two levels are populated by direct formation on grains with rates $k_{f0}$ and $k_{f1}$ or depopulated by  photodissociation  with rates $d_0$ and $d_1$. Other chemical reactions have only a minor impact. The resulting balance equations are

\[
\left( k_{01} \, n\left(\mathrm{H}^+\right) + d_0\right) x_0 - k_{10} \, n\left(\mathrm{H}^+\right) x_1 = k_{f0} \, n\left(\mathrm{H}\right) \, \frac{n_\mathrm{H}}{n\left(\mathrm{H}_2\right)},
\]
\[
\left( k_{10} \, n\left(\mathrm{H}^+\right) + d_1\right) x_1 - k_{01} \, n\left(\mathrm{H}^+\right) x_0 = k_{f1} \, n\left(\mathrm{H}\right) \, \frac{n_\mathrm{H}}{n\left(\mathrm{H}_2\right)}.
\]
This system can be solved for $x_0$ and $x_1$ and leads to

\[
\frac{x_1}{x_0} = \frac{k_{01} \, n\left(\mathrm{H}^+\right) \, k_{f0} + \left( k_{01} \, n\left(\mathrm{H}^+\right) + d_0\right) \, k_{f1}}{\left( k_{10} \, n\left(\mathrm{H}^+\right) + d_1\right) \, k_{f0} + k_{10} \, n\left(\mathrm{H}^+\right) \, k_{f1}}.
\]
In the temperature range from $50$ to $100\mathrm{K}$  appropriate do diffuse and translucent cloud conditions, the Boltzmann factor $\frac{g_1}{g_0} \, \exp\left( -\frac{E_{10}}{T}\right)=  \,9 \, \exp\left( -\frac{170.5}{T}\right) $ varies from $0.5$ to $1.6$. So, $k_{01}$ and $k_{10}$ remain close to one another. Furthermore, the formation rates $k_{f0}$ and $k_{f1}$ are close to one another and simplify. The ratio can be arranged using detailed balance as:

\[
\frac{x_1}{x_0} = \frac{g_1}{g_0} \, \exp\left( -\frac{E_{10}}{T}\right) \, \frac{1 + \frac{d_0}{2 \, k_{01} \, n\left(\mathrm{H}^+\right)}}{1 + \frac{d_1}{2 \, k_{10} \, n\left(\mathrm{H}^+\right)}}
\]
In regions of low radiation field where most diffuse clouds are found the $\mathrm{H}/\mathrm{H}_2$ transition is very close to the edge of the cloud, as shown in \cite{sternberg2014}. Hence, in most of the region that builds $\mathrm{H}_2$ column density the dissociation rates $d_0$ and $d_1$ are about $4$ orders of magnitude lower than the products $k_{10} \, n\left(\mathrm{H}^+\right)$ and $k_{01} \, n\left(\mathrm{H}^+\right)$. Thus, the correction factor coming from the chemistry is very close to $1$ and the ratio $\frac{x_1}{x_0}$ gives a very good measure of the kinetic temperature.

\section{Computation of the electronic fraction}
\label{ab:xe}

Ionization balance can be solved  analytically for diffuse cloud conditions. Due to ultraviolet photons, all metals with an ionization threshold below $13.6\,\mathrm{eV}$ are ionized, providing a minimal electronic abundance of $\delta_M\,n_\mathrm{H}$, where $\delta_M$ is the fraction of relevant metals (mostly $\mathrm{C}$ and $\mathrm{S}$, with traces of $\mathrm{Si}$ and other heavier species). In the following, we take $\delta_M = 1.55\,10^{-4}$.

Additional electrons come from $\mathrm{H}^+$ and $\mathrm{He}^+$ resulting from the balance between ionization via cosmic rays and recombination with electrons and grains.
We follow here for the most part the presentation of \cite{draine2011}, Section 13.6, extended to include $\mathrm{He}$. The balance equations are

\[
\zeta_{\mathrm{H}} \, n\left(\mathrm{H}\right) = \alpha_{rr}\left(\mathrm{H}^{+}\right) \, n\left(\mathrm{H}^{+}\right) \, n\left(e^{-}\right) + \alpha_{gr}\left(\mathrm{H}^{+}\right) \, n_{\mathrm{H}} \, n\left(\mathrm{H}^{+}\right),
\]
\[
\zeta_{\mathrm{He}} \, n\left(\mathrm{He}\right) = \alpha_{rr}\left(\mathrm{He}^{+}\right) \, n\left(\mathrm{He}^{+}\right) \, n\left(e^{-}\right) + \alpha_{gr}\left(\mathrm{He}^{+}\right) \, n_{\mathrm{H}} \, n\left(\mathrm{He}^{+}\right).
\]
Where $\alpha_{rr}$ is the radiative recombination rate, and $\alpha_{gr}$ the rate of recombination on grains. $\zeta_{\mathrm{H}}$ and $\zeta_{\mathrm{He}}$ are the cosmic ray ionization rates of $\mathrm{H}$ and $\mathrm{He}$, respectively. With respect to $\mathrm{H}_2$, we use $\zeta_{\mathrm{H}}= 0.77\,\zeta$, including secondary ionization, and $\zeta_{\mathrm{He}}= 0.5\,\zeta$. The total abundance of electrons is
\[
n\left(e^{-}\right) = \delta_{\mathrm{M}} \, n_{\mathrm{H}} + n\left(\mathrm{H}^{+}\right) + n\left(\mathrm{He}^{+}\right).
\]
For all abundances, we write $x\left(\mathrm{X}\right) = n\left(\mathrm{X}\right) \, n_{\mathrm{H}}$. We take $x\left(\mathrm{He}\right) = \delta_{\mathrm{He}}$, with $\delta_{\mathrm{He}} = 0.1$ and compute $\mathrm{H}$ abundance from the molecular fraction $f_m$

\[
n\left(\mathrm{H}\right) = \left(1 - f_{m} - x\left(\mathrm{H}^{+}\right)\right) \, n_{\mathrm{H}}.
\]
The resulting system of two equations can be written in a more compact form by using $X$ for hydrogen and $Y$ for helium

\[
\alpha_{rr}^{X}\,n_{\mathrm{H}}\,X^{2}+\alpha_{rr}^{X}\,n_{\mathrm{H}}\,X\,Y+\left(\zeta_{\mathrm{H}}+\alpha_{rr}^{X}\,\delta_{\mathrm{M}}\,n_{\mathrm{H}}+\alpha_{gr}^{X}\,n_{\mathrm{H}}\right)\,X=\zeta_{\mathrm{H}}\,\left(1-f_{m}\right),
\]
\[
\alpha_{rr}^{Y}\,n_{\mathrm{H}}\,X\,Y+\alpha_{rr}^{Y}\,n_{\mathrm{H}}\,Y^{2}+\left(\alpha_{rr}^{Y}\,\delta_{\mathrm{M}}+\alpha_{gr}^{Y}\right)\,n_{\mathrm{H}}\,Y=\zeta_{\mathrm{He}}\,\delta_{\mathrm{He}}.
\]
This system is easily solved using a Newton-Raphson scheme once the rates are known.

Radiative electronic recombination rates are taken from \cite{badnell2006}, the relevant coefficients are given in Table\,\ref{Badnell_coef}

\[
\alpha_{rr} = A \times \left[ \sqrt{\frac{T}{T_0}} \left( 1 + \sqrt{\frac{T}{T_0}} \right)^{1 - BB} \times \left( 1 + \sqrt{\frac{T}{T_1}} \right)^{1 + BB}
\right]
\]
with
\[
BB = B + C \, \exp\left( -\frac{T_2}{T}\right).
\]

\begin{table*}[b]
\caption{Electronic recombination rates}
\label{Badnell_coef}
\begin{center}
\begin{tabular}{ccccccc}
\hline
\hline
{Coefficient} & $A$ & $B$ & $T_0$ & $T_1$ & $C$ & $T_2$\tabularnewline 
\hline
$\mathrm{H}^+$ & $8.32\,10^{-11}$ & $0.7472$ & $2.96$ & $7.0\,10^{5}$ & $0$ & $0$\tabularnewline
$\mathrm{He}^+$ & $5.23\,10^{-11}$ & $0.6988$ & $7.3$ & $4.48\,10^{6}$ & $0.0829$ & $1.68\,10^{5}$\tabularnewline
\hline
\end{tabular}
\par\end{center}
\end{table*}

Electronic recombination on grains, comes from \cite{Weingartner2001} 

\[
\alpha_{gr}\left(X^{+}, \frac{G_{0}}{n\left(e^{-}\right)}, T\right) = \frac{10^{-14} \, C_{0}}{1+C_{1} \, \psi^{C_{2}} \, \left(1 + C_{3} \, T^{C_{4}} \, \psi^{-C_{5}-C_{6} \, \ln T}\right)}
\]
with
\[
\psi = \frac{G_{0} \, \sqrt{T}}{n\left(e^{-}\right)}.
\]
Here, $G_0$ is the Inter Stellar Radiation Field (ISRF) intensity in units of Draine's ISRF. Coefficients $C_0$ to $C_6$ are given in Table\,(\ref{tab:GR_coef}).

\begin{table*}[h]
\caption{Grain recombination rates}
\label{tab:GR_coef}
\begin{center}
\begin{tabular}{cccccccc}
\hline
\hline
{Coefficient} & $C_{0}$ & $C_{1}$ & $C_{2}$ & $C_{3}$ & $C_{4}$ & $C_{5}$ & $C_{6}$\tabularnewline 
\hline
$\mathrm{H}^+$ & $12.25$ & $8.074\,10^{-6}$ & $1.378$ & $5.087\,10^{2}$ & $1.586\,10^{-2}$ & $0.4723$ & $1.102\,10^{-5}$\tabularnewline
$\mathrm{He}^+$ & $5.572$ & $3.185\,10^{-7}$ & $1.512$ & $5.115\,10^{3}$ & $3.902\,10^{-7}$ & $0.4956$ & $5.494\,10^{-7}$\tabularnewline
\hline
\end{tabular}
\par\end{center}
\end{table*}

\section{Fortran code}

The code solving for $\mathrm{H}_3^+$ populations from Eq.\,\ref{eq:Det_Bal} is available at 
\href{http://pdr.obspm.fr/index.html}{Meudon ISM Services Platform}. Compilation requires a modern Fortran 90 compiler (gfortran will do) and access to the LAPACK library. The later is usually provided with all standard compilers. Otherwhile, it is self contained.

The code takes very few input parameters: $\left(n_{\mathrm{H}},\,T,\,f_{m},\,I_{m},\,\zeta\right)$ and the number of levels used, from the command line or redirection of a small input file. It uses the latest data available, as described in this paper. Comments in the source file, coupled to this paper, should be enough for easy use and adaptation.

\section*{Disclosure statement}

No conflict of interest.

\section*{Funding}

ER, JLB \& FLP were supported in part by the Programme National “Physique et Chimie du Milieu Interstellaire” (PCMI) of CNRS/INSU with INC/INP, co-funded by CEA and CNES. FK, AO \& HK acknowledge financial support by the Max Planck Society.

\pagebreak

\bibliographystyle{tfo}

\end{document}